\documentclass{article}
\usepackage{frascatiphys,here,graphicx,subfigure}
\begin{document}

\title{ISOSPIN EFFECTS ON MESON PRODUCTION IN RELATIVISTIC HEAVY ION 
COLLISIONS}

\author
{ M. Di Toro, M. Colonna, G. Ferini, V. Greco, J. Rizzo
\\
{\em LNS-INFN 
and Physics-Astronomy Dept., Univ. of Catania, Italy}
\\
V. Baran
\\
{\em NIPNE-HH and Bucharest University, Romania}
\\
T. Gaitanos
\\
{\em Institut f\"ur Theoretische Physik, Universit\"at Giessen, 
 Germany}
\\
Liu Bo
\\
{\em IHEP, Beijing, China}
\\
G. Lalazissis, V. Prassa
\\
{\em Dept. of Theoretical Physics, Aristotle
University, Thessaloniki, Greece}
\\
H. H. Wolter
\\
{\em Dept. f\"ur Physik, 
Universit\"at M\"unchen, Germany}
}

\maketitle

\baselineskip=11.6pt

\begin{abstract}
We show that the phenomenology of isospin effects on heavy ion reactions
at intermediate energies (few $AGeV$ range) is extremely rich and can allow
a ``direct'' study of the covariant structure of the isovector interaction
in a high density hadron medium. We work within a relativistic transport frame,
beyond a cascade picture, 
 consistently derived from effective Lagrangians, where isospin effects
are accounted for in the mean field and collision terms.
We show that rather sensitive observables are provided by the 
pion/kaon production ($\pi^-/\pi^+$, 
$K^0/K^+$ yields). Relevant non-equilibrium effects
are stressed.
The possibility of the transition to a mixed hadron-quark phase, 
at high baryon and isospin density, is finally suggested. Some signatures
could come from an expected ``neutron trapping'' effect.

\end{abstract}

\baselineskip=14pt


\vskip -1.0cm
\section{Introduction}

Recently the development of new heavy ion facilities (radioactive beams) 
has driven 
the interest on the dynamical behaviour of asymmetric matter, 
\cite{baranPR}. 
Here we focus our attention on relativistic heavy ion collisions, that
provide a unique terrestrial opportunity to probe the in-medium nuclear
interaction in high density and high momentum regions. 
An effective Lagrangian approach to the hadron interacting system is
extended to the isospin degree of freedom: within the same frame equilibrium
properties ($EoS$, \cite{qhd}) and transport dynamics 
can be consistently derived.

Within a covariant picture of the nuclear mean field, 
 for the description of the symmetry energy at saturation
($a_{4}$ parameter of the Weizs\"{a}ecker mass formula)
(a) only the Lorentz vector $\rho$ mesonic field, 
and (b) both, the vector $\rho$ (repulsive) and  scalar 
$\delta$ (attractive) effective 
fields \cite{liu,gait04} can be included. 
In the latter case the competition between scalar and vector fields leads
to a stiffer symmetry term at high density \cite{liu,baranPR}. We present
here observable effects in the dynamics of heavy ion 
collisions. 
We focus our attention on the isospin content of meson 
production.
We finally show that in the compression stage of isospin asymmetric collisions
we can even enter a mixed deconfined phase.
\vskip -1.0cm
\section{Relativistic Transport}
The starting point is
a simple phenomenological version of the Non-Linear (with respect to the 
iso-scalar, Lorentz scalar $\sigma$ field) effective nucleon-boson 
field theory,  
the Quantum-Hadro-Dynamics \cite{qhd}. 
According to this picture 
the presence of the hadronic medium leads to effective masses and 
momenta $M^{*}=M+\Sigma_{s}$,   
 $k^{*\mu}=k^{\mu}-\Sigma^{\mu}$, with
$\Sigma_{s},~\Sigma^{\mu}$
 scalar and vector self-energies. 
For asymmetric matter the self-energies are different for protons and 
neutrons, depending on the isovector meson contributions. 
We will call the 
corresponding models as $NL\rho$ and $NL\rho\delta$, respectively, and
just $NL$ the case without isovector interactions. 
For the more general $NL\rho\delta$ case  
the self-energies 
of protons and neutrons read:
\begin{eqnarray}
\Sigma_{s}(p,n) = - f_{\sigma}\sigma(\rho_{s}) \pm f_{\delta}\rho_{s3}, 
\nonumber \\
\Sigma^{\mu}(p,n) = f_{\omega}j^{\mu} \mp f_{\rho}j^{\mu}_{3},
\label{selfen}
\end{eqnarray}
(upper signs for neutrons), where $\rho_{s}=\rho_{sp}+\rho_{sn},~
j^{\alpha}=j^{\alpha}_{p}+j^{\alpha}_{n},\rho_{s3}=\rho_{sp}-\rho_{sn},
~j^{\alpha}_{3}=j^{\alpha}_{p}-j^{\alpha}_{n}$ are the total and 
isospin scalar 
densities and currents and $f_{\sigma,\omega,\rho,\delta}$  are the coupling 
constants of the various 
mesonic fields. 
$\sigma(\rho_{s})$ is the solution of the non linear 
equation for the $\sigma$ field \cite{liu,baranPR}.

For the description of heavy ion collisions we solve
the covariant transport equation of the Boltzmann type 
within the 
Relativistic Landau
Vlasov ($RLV$) method, using phase-space Gaussian test particles 
\cite{FuchsNPA589},
and applying
a Monte-Carlo procedure for the hard hadron collisions.
The collision term includes elastic and inelastic processes involving
the production/absorption of the $\Delta(1232 MeV)$ and $N^{*}(1440
MeV)$ resonances as well as their decays into pion channels,
 \cite{FeriniNPA762}.
\vskip -1.0cm
\section{Isospin effects on pion and kaon production at intermediate 
energies} 
Kaon production has been proven to be a reliable observable for the
high density $EoS$ in the isoscalar sector 
\cite{FuchsPPNP56,HartPRL96}
Here we show that the $K^{0,+}$
production (in particular the $K^0/K^+$ yield ratio) can be also used to 
probe the isovector part of the $EoS$,
\cite{Fer06,Pra07}.

Using our $RMF$ transport approach  we analyze 
pion and kaon production in central $^{197}Au+^{197}Au$ collisions in 
the $0.8-1.8~AGeV$
 beam 
energy range, comparing models giving the same ``soft'' $EoS$ for symmetric 
matter and with different effective field choices for 
$E_{sym}$. We will use three Lagrangians with constant 
nucleon-meson 
couplings ($NL...$ type, see before) and one with density
dependent couplings ($DDF$, see \cite{gait04}), recently suggested 
for better nucleonic properties of neutron stars \cite{Klahn06,Liubo07}.

\begin{figure}[t] 
\begin{center}
\includegraphics[scale=0.27]{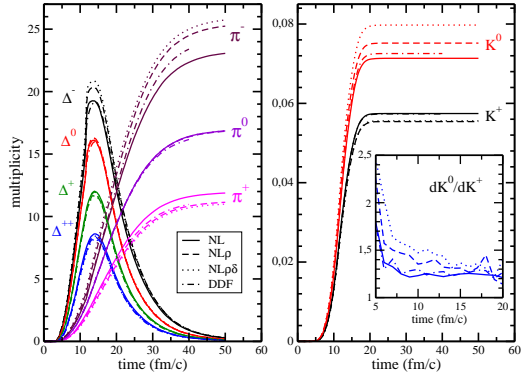} 
\vskip -0.3cm
\caption{\small{Time evolution of the $\Delta^{\pm,0,++}$ resonances
and pions $\pi^{\pm,0}$ 
(left),  and  kaons ($K^{+,0}$
 (right) for a central ($b=0$ fm impact parameter)  
Au+Au collision at 1 AGeV incident energy. Transport calculation using the  
$NL, NL\rho, NL\rho\delta$ and $DDF$ models for the iso-vector part of the  
nuclear $EoS$ are shown. The inset contains the differential $K^0/K^+$  ratio
as a function of the kaon emission time.  
}}
\vskip -0.7cm
\label{kaon1} 
\end{center}
\end{figure}

Fig. \ref{kaon1} reports  the temporal evolution of $\Delta^{\pm,0,++}$  
resonances, pions ($\pi^{\pm,0}$) and kaons ($K^{+,0}$)  
for central Au+Au collisions at $1AGeV$.
It is clear that, while the pion yield freezes out at times of the order of 
$50 fm/c$, i.e. at the final stage of the reaction (and at low densities),
kaon production occur within the very early (compression) stage,
 and the yield saturates at around $20 fm/c$. 
From Fig. \ref{kaon1} we see that the pion results are  
weakly dependent on the  
isospin part of the nuclear mean field.
However, a slight increase (decrease) in the $\pi^{-}$ ($\pi^{+}$) 
multiplicity is observed when going from the $NL$ (or $DDF$) to the 
$NL\rho$ and then to
the $NL\rho\delta$ model, i.e. increasing the vector contribution $f_\rho$
in the isovector channel. This trend is 
more pronounced for kaons, see the
right panel, due to the high density selection of the source and the
proximity to the production threshold. Consistently, as shown in the
insert, larger effects are expected for early emitted kaons, reflecting the 
early $N/Z$ of the system. 

When isovector fields are included the symmetry potential energy in 
neutron-rich matter is repulsive for neutrons and attractive for protons.
In a $HIC$ this leads to a fast, pre-equilibrium, emission of neutrons.
 Such a $mean~field$ mechanism, often referred to as isospin fractionation
\cite{baranPR}, is responsible for a reduction of the neutron
to proton ratio during the high density phase, with direct consequences
on particle production in inelastic $NN$ collisions.

$Threshold$ effects represent a more subtle point. The energy 
conservation in
a hadron collision in general has to be formulated in terms of the canonical
momenta, i.e. for a reaction $1+2 \rightarrow 3+4$ as
$
s_{in} = (k_1^\mu + k_2^\mu)^2 = (k_3^\mu + k_4^\mu)^2 = s_{out}.
$
Since hadrons are propagating with effective (kinetic) momenta and masses,
 an equivalent relation should be formulated starting from the effective
in-medium quantities $k^{*\mu}=k^\mu-\Sigma^\mu$ and $m^*=m+\Sigma_s$, where
$\Sigma_s$ and $\Sigma^\mu$ are the scalar and vector self-energies,
Eqs.(\ref{selfen}).
The self-energy contributions will influence the particle production at the
level of thresholds as well as of the phase space available in the final 
channel. In fact the {\it threshold} effect is dominant and consequently the
results are nicely sensitive to the covariant structure of the isovector
fields.
At each beam energy we see an
increase of the $\pi^-/\pi^+$ and 
$K^{0}/K^{+}$ 
yield ratios with the models
$NL \rightarrow DDF \rightarrow NL\rho \rightarrow NL\rho\delta$. 
The effect is larger for the $K^{0}/K^{+}$ compared to the $\pi^-/\pi^+$
ratio. This is due to the subthreshold production and to the fact that
the isospin effect enters twice in the two-step production of kaons, see
\cite{Fer06}. 
Interestingly the Iso-$EoS$ effect for pions is increasing at lower energies,
when approaching the production threshold.


We have to note that in a previous study of kaon production in excited nuclear
matter the dependence of the $K^{0}/K^{+}$ yield ratio on the effective
isovector interaction appears much larger (see Fig.8 of 
ref.\cite{FeriniNPA762}).
The point is that in the non-equilibrium case of a heavy ion collision
the asymmetry of the source where kaons are produced is in fact reduced
by the $n \rightarrow p$ ``transformation'', due to the favored 
$nn \rightarrow p\Delta^-$ processes. This effect is almost absent at 
equilibrium due to the inverse transitions, see Fig.3 of 
ref.\cite{FeriniNPA762}. Moreover in infinite nuclear matter even the fast
neutron emission is not present. 
This result clearly shows that chemical equilibrium models can lead to
uncorrect results when used for transient states of an $open$ system.
\vskip -1.0cm
\section{Testing Deconfinement at High Isospin Density}
The hadronic matter is expected to undergo a phase transition 
to a deconfined phase of quarks and gluons at large densities 
and/or high temperatures. On very general grounds,
the transition's critical densities are expected to depend
on the isospin of the system, but no experimental tests of this 
dependence have been performed so far.
\begin{figure}
\begin{center}
\includegraphics[scale=0.28]{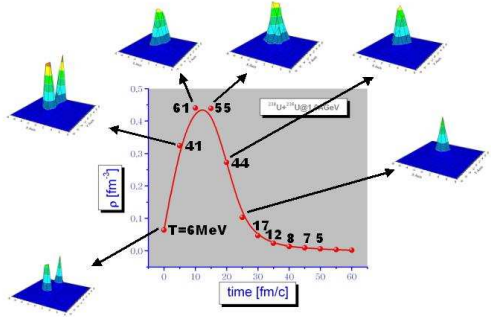}
\vskip -0.5cm
\caption{
\small{$^{238}U+^{238}U$, $1~AGeV$, semicentral. Correlation between 
density, 
temperature, momentum
thermalization inside a cubic cell, 2.5 $fm$ wide, in the center
of mass of the system.  
}}
\label{figUU}
\vskip -0.7cm
\end{center}
\end{figure}
In order to check the possibility of observing some precursor signals
of a new physics even in collisions of stable nuclei at
intermediate energies we have performed some event simulations for the
collision of very heavy, neutron-rich, elements. We have chosen the
reaction $^{238}U+^{238}U$ (average proton fraction $Z/A=0.39$) at
$1~AGeV$ and semicentral impact parameter $b=7~fm$ just to increase
the neutron excess in the interacting region. 
In  Fig.~\ref{figUU} we report the evolution of momentum distribution
and baryon density in a space cell located in the c.m. of the system.
We see that after about $10~fm/c$ a local
equilibration is achieved.  We have a unique Fermi distribution and
from a simple fit we can evaluate the local temperature 
(black numbers in MeV).
We note that a rather exotic nuclear matter is formed in a transient
time of the order of $10~fm/c$, with baryon density around $3-4\rho_0$,
temperature $50-60~MeV$, energy density $500~MeV~fm^{-3}$ and proton
fraction between $0.35$ and $0.40$, likely inside the estimated mixed 
phase region.

In fact we can study the isospin dependence of the transition densities
\cite{deconf06}.
The structure of the mixed phase is obtained by
imposing the Gibbs conditions \cite{Landaustat} for
chemical potentials and pressure and by requiring
the conservation of the total baryon and isospin densities
\begin{eqnarray}\label{gibbs}
&&\mu_B^{(H)} = \mu_B^{(Q)}\, ,~~  
\mu_3^{(H)} = \mu_3^{(Q)} \, ,  \nonumber \\
&&P^{(H)}(T,\mu_{B,3}^{(H)}) = P^{(Q)} (T,\mu_{B,3}^{(Q)})\, ,\nonumber \\
&&\rho_B=(1-\chi)\rho_B^H+\chi\rho_B^Q \, , \nonumber \\
&&\rho_3=(1-\chi)\rho_3^H+\chi\rho_3^Q\, , 
\end{eqnarray}
where $\chi$ is the fraction of quark matter in the mixed phase.
\begin{figure}
\begin{center}
\includegraphics[angle=+90,scale=0.28]{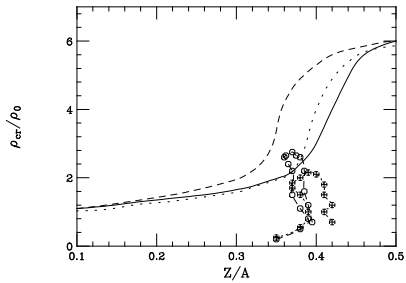}
\vskip -0.3cm
\caption{
\small{Variation of the transition density with proton fraction for various
hadronic $EoS$ parameterizations. Dotted line: $GM3$ $RMF$-model
\cite{GlendenningPRL18};
 dashed line: $NL\rho$ ; solid line: $NL\rho\delta$ . 
For the quark $EoS$: $MIT$ bag model with
$B^{1/4}$=150 $MeV$.
The points represent the path followed
in the interaction zone during a semi-central $^{132}$Sn+$^{132}$Sn
collision at $1~AGeV$ (circles) and at $300~AMeV$ (crosses). 
}}
\vskip -1.0cm
\label{rhodelta}
\end{center}
\end{figure}
In this way we get the $binodal$ surface which gives the phase coexistence 
region
in the $(T,\rho_B,\rho_3)$ space.
For a fixed value of the
conserved charge $\rho_3$ 
 we will study the boundaries of the mixed phase
region in the $(T,\rho_B)$ plane. 
In the hadronic phase the charge chemical potential is given by
$
\mu_3 = 2 E_{sym}(\rho_B) \frac{\rho_3}{\rho_B}\, .
$ 
Thus, we expect critical densities
rather sensitive to the isovector channel in the hadronic $EoS$.

In Fig.~\ref{rhodelta}  we show the crossing
density $\rho_{cr}$ separating nuclear matter from the quark-nucleon
mixed phase, as a function of the proton fraction $Z/A$.  
We can see the effect of the
$\delta$-coupling towards an $earlier$ crossing due to the larger
symmetry repulsion at high baryon densities.
In the same figure we report the paths in the $(\rho,Z/A)$
plane followed in the c.m. region during the collision of the n-rich
 $^{132}$Sn+$^{132}$Sn system, at different energies. At
$300~AMeV$ we are just reaching the border of the mixed phase, and we are
well inside it at $1~AGeV$. 
We expect a {\it neutron trapping}
effect, supported by statistical fluctuations as well as by a 
symmetry energy difference in the
two phases.
In fact while in the hadron phase we have a large neutron
potential repulsion (in particular in the $NL\rho\delta$ case), in the
quark phase we only have the much smaller kinetic contribution.
Observables related to such neutron ``trapping'' could be an
inversion in the trend of the formation of neutron rich fragments
and/or of the $\pi^-/\pi^+$, $K^0/K^+$ yield ratios for reaction
products coming from high density regions, i.e. with large transverse
momenta.  
\vskip -1.5cm
\section{Perspectives}
We have shown that meson production in n-rich heavy ions collisions
at intermediate energies
can bring new information on the isovector part of the in-medium interaction
at high baryon densities.
Important non-equilibrium effects for particle production are stressed.
Finally the possibility of observing
precursor signals of the phase transition to a mixed hadron-quark matter
at high baryon density is suggested.
%

\noindent
{\it Acknowledgements}.
We warmly thanks A.Drago and A.Lavagno for the intense 
collaboration on the
mixed hadron-quark phase transition at high baryon and isospin density.



\end{document}